\documentclass[twocolumn,showpacs]{revtex4}
\usepackage{amsmath,amsthm,amssymb,amscd,float,accents}
\usepackage{graphicx}
\usepackage[T1]{fontenc}
\usepackage{lmodern}
\newcommand{\al}{\alpha}

\newcommand{\de}{\delta}

\newcommand{\vep}{\varepsilon}
\newcommand{\ga}{\gamma}

\newcommand{\la}{\lambda}
\newcommand{\om}{\omega}
\newcommand{\si}{\sigma}

\newcommand{\vp}{\varphi}
\newcommand{\ze}{\zeta}
\newcommand{\La}{\Lambda}

\newcommand{\bde}{\boldsymbol{\delta}}

\newcommand{\bk}{\mathbf{k}}

\newcommand{\bn}{\mathbf{n}}

\newcommand{\CC}{{\mathbb C}}

\newcommand{\ZZ}{{\mathbb Z}}
\newcommand{\cF}{{\mathcal F}}

\newcommand{\cP}{{\mathcal P}}

\newcommand{\id}{1\hspace{-.25em}{\rm I}}

\def\ket#1{|#1\rangle}

\newcommand{\ms}{\mspace{1mu}}

\renewcommand{\le}{\leqslant}
\renewcommand{\ge}{\geqslant}
\def\bbuildrel#1_#2^#3{\mathrel{\mathop{\kern0pt #1}\limits_{#2}^{#3}}}
\newcommand{\tends}[1]{\bbuildrel{\hbox to 2em{\rightarrowfill}}_{#1}^{}}
\newcommand{\erf}{\operatorname{erf}}

\newcommand{\tr}{\operatorname{tr}}

\newcommand{\diag}{\operatorname{diag}}

\newcommand{\iu}{i}
\newcommand{\e}{e}

\newcounter{ex}

\begin{document}
\title{On the level density of spin chains of Haldane--Shastry type}
\author{Alberto \surname{Enciso}}%
\email{aenciso@fis.ucm.es}%
\author{Federico \surname{Finkel}}%
\email{ffinkel@fis.ucm.es} \author{Artemio \surname{Gonz\'alez-L\'opez}}%
\email[Corresponding author. Electronic address: ]{artemio@fis.ucm.es}%
\affiliation{Departamento de F\'\i sica Te\'orica II, Universidad Complutense, 28040 Madrid, Spain}
\date{May 18, 2010}
\begin{abstract}
  We provide a rigorous proof of the fact that the level density of all $\mathrm{su}(m)$ spin
  chains of Haldane--Shastry type associated with the $A_{N-1}$ root system approaches a Gaussian
  distribution as the number of spins $N$ tends to infinity. Our approach is based on the study of
  the large $N$ limit of the characteristic function of the level density, using the description
  of the spectrum in terms of motifs and the asymptotic behavior of the dispersion relation.
\end{abstract}
\pacs{02.50.Cw, 05.30.-d, 75.10.Pq}
\maketitle
%
%%%%%%%%%%%%%%%%%%%%%%%%%%%%%%%%% 80 characters %%%%%%%%%%%%%%%%%%%%%%%%%%%%%%%%%

A well-known conjecture on spin chains of Haldane--Shastry (HS)
type~\cite{Ha88,Sh88,Po93,FI94,BPS95,YT96,FGGRZ03,EFGR05,BFG09,BFG09pre} states that their level
density becomes Gaussian as the number of sites tends to infinity. Although this conjecture has
been numerically verified for all chains of HS type whose spectrum has been computed in closed
form~\cite{EFGR05,FG05,BB06,BFGR08,BFGR08epl,BFGR09,BFG09,BB09,BFGR10}, a rigorous proof thereof
is lacking, except in a few exceptional cases in which the partition function
factorizes~\cite{EFG09}. In this paper we settle the conjecture in the affirmative for spin chains
of HS type associated with the $A_{N-1}$ root system, which is both the simplest and the most
studied case.

Our result has implications in connection with two fundamental conjectures in the theory of
quantum chaos that we shall now discuss. The first of these conjectures, due to Berry and
Tabor~\cite{BT77}, asserts that the probability density of spacings between consecutive levels in
the spectrum of a quantum system whose classical analog is integrable follows Poisson's law
$p(s)=\e^s$. The second conjecture, formulated by Bohigas, Giannoni and Schmidt~\cite{BGS84},
posits that for a fully chaotic quantum system this density is instead given by Wigner's surmise
$p(s)=(\pi s/2)\exp(-\pi s^2/4)$, characteristic of the Gaussian orthogonal ensemble (GOE) in
random matrix theory~\cite{Me04}. It is important to bear in mind that in this context the term
``spectrum'' refers to what is known as the \emph{unfolded} spectrum, which by construction has an
approximately uniform level density. The energies of this unfolded spectrum are obtained from the
``raw'' energies $E_i$ through the mapping $E_i\mapsto \eta_i\equiv\eta(E_i)$, where $\eta$ is a
continuous approximation to the cumulative level density. Thus, the knowledge of this continuous
approximation is essential for testing the latter conjectures. It turns out that in all spin
chains of HS type studied so far, if one assumes that the continuous part of the cumulative level
density is Gaussian, i.e.,
\[
\eta(E)=\frac1{\sqrt{2\pi}\,\si}\int_{-\infty}^E\e^{-\frac{(E'-\mu)^2}{2\si^2}} dE'
  = \frac12\bigg[1+\erf\bigg(\frac{E-\mu}{\sqrt2\,\si}\bigg)\bigg]
\]
(where $\mu$ and $\si$ are the mean and standard deviation of the spectrum), the spacings density
follows a characteristic distribution which is neither of Poisson's nor of Wigner's
type~\cite{FG05,BB06,BFGR08,BFGR08epl,BFGR09,BFG09,BB09,BFGR10}.

Let us briefly recall the definition of the three spin chains we shall deal with in what follows.
The Hamiltonian of the original $\mathrm{su}(m)$ Haldane--Shastry chain is defined as
\begin{equation}\label{HS}
H=\frac12\sum_{i<j}\frac{1-\vep S_{ij}}{\sin^2(\xi_i-\xi_j)}\,,\qquad \xi_k=\frac{k\pi}N\,,
\end{equation}
where (as always hereafter, unless otherwise stated) the sum runs from $1$ to the number of spins
$N$, and $\vep=1$ (resp.~$\vep=-1$) for the ferromagnetic (resp. antiferromagnetic) chain. The
operator $S_{ij}$ permutes the $i$th and $j$th spins, i.e., its action on an element
$\ket{s_1,\dots,s_i,\dots,s_j,\dots,s_N}$ ($s_k\in\{1,\dots,m\}$) of the canonical spin basis is
given by
\[
S_{ij}\ket{\dots,s_i,\dots,s_j,\dots}=\ket{\dots,s_j,\dots,s_i,\dots}\,.
\]
The permutation operators $S_{ij}$ can be expressed in terms of the generators $t^a_k$ of the
fundamental representation of $\mathrm{su}(m)$ at the $k$-th site as
\[
S_{ij}=2\sum_{a=1}^{m^2-1}t_i^a t_j^a+\frac1m\,,
\]
with the normalization $\tr(t^a_kt^b_k)=\frac12\de^{ab}$. The chain~\eqref{HS} is intimately
connected with the Hubbard model. For instance, it can be obtained from the one-dimensional
Hubbard model with long range hopping introduced in Ref.~\cite{GR92} when the on-site interaction
tends to infinity and the sites are half-filled. The rational version of the HS chain~\eqref{HS}
was subsequently introduced by Polychronakos~\cite{Po93} and Frahm~\cite{Fr93}. The Hamiltonian of
the Polychronakos--Frahm (PF) chain can be taken as
\begin{equation}\label{PF}
H=\sum_{i<j}\frac{1-\vep S_{ij}}{(\xi_i-\xi_j)^2}\,,
\end{equation}
where the chain sites $\xi_k$ are no longer equidistant, but are given by the zeros of the Hermite
polynomial of degree $N$. Finally, the hyperbolic version of the HS chain, known as the
Frahm--Inozemstsev (FI) chain~\cite{FI94}, is defined by the Hamiltonian
\begin{equation}\label{FI}
  H=\frac12\,\sum_{i<j}\frac{1-\vep S_{ij}}{\sinh^2(\xi_i-\xi_j)}\,.
\end{equation}
The chain sites in this case are given by $\xi_i=\frac 12\log\ze_i$, where $\ze_i$ is the $i$th
zero of the Laguerre polynomial $L_N^{\al-1}$ with $\al>0$. In particular, unlike the previous two
chains, the sites of the FI chain depend on an essential parameter.

Each of the chains~\eqref{HS}--\eqref{FI} can be obtained from a corresponding spin dynamical
model of Calogero--Sutherland type~\cite{Ca71,Su71,Su72} by applying the so-called freezing
trick~\cite{Po93}. As first shown by Polychronakos~\cite{Po94}, this connection can be exploited
to derive closed-form expressions for the partition functions of the above chains. It turns out
that these expressions can be rewritten in a remarkable unified way
as~\cite{FG05,BBS08,BFGR08epl,BFGR10}
\begin{equation}
  \label{Z}
  Z(q) = \sum_{\bk\in\cP_N}\prod_{i=1}^rd(k_i)\cdot
  q^{\sum_{i=1}^{r-1}\cF(K_i)}
  \prod_{i=1}^{N-r} (1-q^{\cF(K'_i)})\,,
\end{equation}
where $q\equiv\e^{-1/(k_{\mathrm B}T)}$, $\bk\equiv (k_1,\dots,k_r)$ is an element of the set
$\cP_N$ of partitions of $N$ with order taken into account, and the spin degeneracy factor
$d(k_i)$ is given by
\[
d(k_i)=
\begin{cases}
    \binom{m+k_i-1}{k_i}\,,\quad &\vep=1\\[2mm]
  \binom m{k_i}\,,\quad &\vep=-1\,.
\end{cases}
\]
The numbers $K_i$ in Eq.~\eqref{Z} are defined as $K_i=\sum_{j=1}^ik_j\in\{1,\dots,N-1\}$, and
$\{K'_1,\dots,K'_{N-r}\}=\{1,\dots,N-1\}\setminus\{K_1,\dots,K_{r-1}\}$. The partition function
$Z$ depends on the chain under consideration only through its \emph{dispersion relation} $\cF(i)$,
given by
\begin{equation}\label{cF}
\cF(i)=
\begin{cases}
  i(N-i)\,,& \text{for the HS chain}\\
  i\,,& \text{for the PF chain}\\
  i(\al+i-1)\,,\quad & \text{for the FI chain}\,.
\end{cases}
\end{equation}
Using Eq.~\eqref{Z}, Basu-Mallick et al.~\cite{BBHS07,BBS08} derived a simple set of rules for
generating the spectrum of the chains~\eqref{HS}--\eqref{FI} in terms of Young tableaux of certain
irreducible representations of the Yangian $Y\big(\mathrm{gl}(m)\big)$. It can be easily shown
that these rules are equivalent to the explicit formula
\begin{equation}\label{energies}
E_{\bn}=\sum_{i=1}^{N-1}\de(n_i,n_{i+1})\,\cF(i)\,,\qquad \bn\equiv(n_1,\dots,n_N)\,,
\end{equation}
where the quantum numbers $n_i$ independently take the values
$1,\dots,m$. As to the function $\de$, it is given by
\begin{equation}\label{dejk}
\de(j,k)=
\begin{cases}
  1\,, \quad &j<k\\[1mm]
  0\,,& j\ge k\,,
\end{cases}
\end{equation}
in the ferromagnetic case, whereas in the antiferromagnetic one it suffices to exchange $0$ and
$1$ in Eq.~\eqref{dejk}. The vectors $\bde(\bn)\in\{0,1\}^{N-1}$ with components
$\de_i(\bn)=\de(n_i,n_{i+1})$ are in fact the celebrated {\em motifs} introduced by Haldane et al.
in Ref.~\cite{HHTBP92}. It should be emphasized that the formula~\eqref{energies} for the energies
is obtained from the partition function and not vice versa, as is usually the case.

Equation~\eqref{energies} shall be our starting point for establishing the asymptotically Gaussian
character of the level density of the chains~\eqref{HS}--\eqref{FI}. In fact, since the sum of the
Hamiltonians of the ferromagnetic and antiferromagnetic chains is a constant, from now on we shall
restrict ourselves to the ferromagnetic case unless otherwise stated. We shall start by deriving a
unified formula for the mean $\mu$ and variance $\si^2$ of the spectrum of the latter chains in
terms of the dispersion function $\cF$. Consider first the mean energy, defined by
\[
\mu=m^{-N}\sum_{n_1,\dots,n_N=1}^m\sum_{i=1}^{N-1}\de(n_i,n_{i+1})\,\cF(i)\,.
\]
By Eq.~\eqref{dejk}, the coefficient of $\cF(i)$ in the previous expression equals $1$ for
$n_{i+1}>n_i$, regardless of the values taken by the $N-2$ remaining quantum numbers $n_k$, and is
$0$ otherwise. Thus
\begin{equation}\label{mu}
\mu=m^{-N}\sum_{i=1}^{N-1}\binom m2m^{N-2}\cF(i)=\frac12\Big(1-\frac 1m\Big)\sum_{i=1}^{N-1}\cF(i)\,.
\end{equation}
Similarly, the variance of the energy is given by $\si^2=\langle E_{\bn}^2\rangle-\mu^2$, where
\begin{multline*}
  \big\langle E_{\bn}^2\big\rangle\equiv m^{-N}\sum_{n_1,\dots,n_N=1}^mE_\bn^2\\
  =m^{-N}\sum_{n_1,\dots,n_N=1}^m\sum_{i,j=1}^{N-1} \de_i(\bn)\de_j(\bn)\cF(i)\cF(j)\,.
\end{multline*}
Taking into account that $\de_i^2=\de_i$ and proceeding as before we easily obtain
\begin{multline*}
  \big\langle E_{\bn}^2\big\rangle=\frac12\Big(1-\frac 1m\Big)\sum_{i=1}^{N-1}\cF(i)^2\\
  +2m^{-N}\sum_{n_1,\dots,n_N=1}^m\,\sum_{\substack{i,j=1\\i<j}}^{N-1}
  \de_i(\bn)\de_j(\bn)\cF(i)\cF(j)\,.
\end{multline*}
If $i<j-1$, the coefficient of $\cF(i)\cF(j)$ in the last sum equals $1$ provided that
$n_i<n_{i+1}$ and $n_j<n_{j+1}$, and is otherwise zero. Likewise, the coefficient of
$\cF(j-1)\cF(j)$ ($j=2,\dots,N-1$) is $1$ if $n_{j-1}<n_j<n_{j+1}$, and vanishes otherwise. Hence
\begin{align*}
  \big\langle E_{\bn}^2\big\rangle={}&\frac12\Big(1-\frac 1m\Big)\sum_{i=1}^{N-1}\cF(i)^2\\
  &+2m^{-N}\sum_{\substack{i,j=1\\i<j-1}}^{N-1}\binom m2^{\!2}m^{N-4}\cF(i)\cF(j)\\
  &+2m^{-N}\sum_{j=2}^{N-1}\binom m3 m^{N-3}\cF(j-1)\cF(j)\,.\\
\end{align*}
Using Eq.~\eqref{mu} for $\mu$, after some straightforward algebra we obtain
\begin{multline}
  \label{si2}
  \si^2
  =\bigg(1-\frac1{m^2}\bigg)\bigg[\frac14\sum_{i=1}^{N-1}\cF(i)^2
  -\frac16\sum_{i=2}^{N-1}\cF(i-1)\cF(i)\bigg].
\end{multline}
Since, up to an additive constant, the energies of the antiferromagnetic
chains~\eqref{HS}--\eqref{FI} differ from those of their ferromagnetic counterparts by a sign
change, it is clear that Eq.~\eqref{si2} is also valid in the antiferromagnetic case. As to
Eq.~\eqref{mu}, using the antiferromagnetic analog of Eq.~\eqref{dejk} and reasoning as before it
is immediate to show that the mean energy of the antiferromagnetic chains is given by
\begin{equation}\label{muAF}
  \mu=\frac12\Big(1+\frac 1m\Big)\sum_{i=1}^{N-1}\cF(i)\,.
\end{equation}
It may be easily verified that the unified expressions~\eqref{si2}-\eqref{muAF} coincide with the
values of $\mu$ and $\si^2$ previously computed on a case by case basis for the
(antiferromagnetic) chains~\eqref{HS}--\eqref{FI}~\cite{FG05,BFGR08epl,BFGR10}.

After these preliminaries, we are now ready to present the main part of our proof. As in our
previous paper~\cite{EFG09}, the proof is based on analyzing the limit as $N$ tends to infinity of
the (normalized) \emph{characteristic function} of the level density, defined by
\begin{align} \hat\vp(t)&=
\Big\langle\e^{\iu t(E_\bn-\mu)/\si}\Big\rangle\equiv
m^{-N}\sum_{n_1,\dots,n_N=1}^m\e^{\iu t(E_\bn-\mu)/\si}\notag\\
\label{charf}
&=m^{-N}\e^{-\iu\mu t/\si}Z\big(\e^{\iu t/\si}\big).
\end{align}
Note that $\hat\vp(t)$ is simply the Fourier transform of the level density, after normalizing the
spectrum to zero mean and unit variance. The importance of the characteristic function in the
present context lies in the following standard result~(see, e.g., Ref.~\cite{Fe71}): in the limit
$N\to\infty$, the level density (normalized to unity) approaches a Gaussian with parameters $\mu$
and $\si$ provided that
\begin{equation} \label{charflim}
  \lim_{N\to\infty}\hat\vp(t)=\e^{-t^2/2}
\end{equation}
for all real $t$. In order to analyze the asymptotic behavior of the characteristic
function~\eqref{charf}, we first rewrite the partition function $Z(q)$ of the
chains~\eqref{HS}--\eqref{FI} using the explicit formula~\eqref{energies} for the energies,
obtaining
\begin{align}\label{ZT}
Z(q)&=\sum_{n_1,\dots,n_N=1}^m \prod_{i=1}^{N-1}q^{\de(n_i,n_{i+1})\cF(i)}\notag\\
&=m^{N-1}\sum_{n,n'=1}^m\big[T_1(q)T_2(q)\cdots T_{N-1}(q)\big]_{nn'}\,,
\end{align}
where the transfer matrix $T_j(q)$ ($j=1,\dots,N-1$) is the $m\times m$ matrix with elements
\[
\big[T_j(q)\big]_{kl}=\frac1m\,q^{\de(k,l)\cF(j)}\,,\qquad 1\le k,l\le m\,.
\vrule width0pt depth 12pt
\]
Note that, since $\de(k,l)$ depends only on the difference $k-l$ (cf.~Eq.~\eqref{dejk}), $T_j(q)$
is a Toeplitz matrix. More precisely, using the explicit definition~\eqref{dejk} we see that
\begin{equation}\label{Tj}
T_j(q)=T\big(\om_j(q)\big)\,,
\end{equation}
where the matrix $T(\om)$ is given by
\begin{equation}
  \label{Tom}
T(\om)=\frac1m
\begin{pmatrix}
  1& \om^m&\cdots & \om^m\\
  \vdots&\ddots&\ddots&\vdots\\
  \vdots&&\ddots&\om^m\\
  1&\cdots&\cdots&1
\end{pmatrix}\,,\qquad\om\in\CC\,,
\end{equation}
and
\begin{equation}\label{omj}
\om_j(q)=q^{\cF(j)/m}\,.
\end{equation}

The $m\times m$ matrix~\eqref{Tom} can be easily diagonalized for arbitrary $m$ and $\om$. Indeed,
consider the vectors $v^k(\om)$ $(k=1,\dots,m)$ with components
\[
v^k_j(\om) = \big[\om\ms\e^{2\pi\iu k/m}\big]^{m-j}\,,\qquad j=1,\dots,m\,;
\]
note, in particular, that $v^m(\om)=(\om^{m-1},\dots,\om,1)$. We then have
\begin{align*}
  m\big[T&(\om)v^k(\om)\big]_j\\ &= \sum_{l=1}^j\big[\om\ms\e^{2\pi\iu
    k/m}\big]^{m-l}+\om^m\sum_{l=j+1}^m\big[\om\ms\e^{2\pi\iu
    k/m}\big]^{m-l}\\&=\sum_{l=1}^j\big[\om\ms\e^{2\pi\iu k/m}\big]^{m-l}
  +\sum_{l=j+1}^m\big[\om\ms\e^{2\pi\iu k/m}\big]^{2m-l}\\
  &=m\ms\la_k(\om)v^k_j(\om)\,,
\end{align*}
where
\begin{equation}
  \label{lak}
  \la_k(\om)=\frac1m\,\sum_{l=0}^{m-1}\big[\om\ms\e^{2\pi\iu k/m}\big]^l\,.
\end{equation}
Thus $v^k(\om)$ is an eigenvector of $T(\om)$ with eigenvalue $\la_k(\om)$. The above result is
valid for arbitrary $\om\in\CC$. When $\om$ is unimodular, the vectors $v^k(\om)/\sqrt{m}$
($k=1,\dots,m$) form an orthonormal basis of $\CC^m$. Indeed, they are clearly of unit length, and
their scalar product is given by
\begin{align*}
  v^k\cdot v^{k'}&=\sum_{l=1}^m\bigg[\overline\om\,\e^{-2\pi\iu k/m}\cdot\om\ms\e^{2\pi\iu
    k'/m}\bigg]^{m-l}\\
  &=\sum_{l=0}^{m-1}\e^{2\pi\iu(k'-k)l/m}=0\,,\qquad k\ne k'\,,
\end{align*}
where we have used the fact that $\overline\om=\om^{-1}$. In other words, when $|\om|=1$ we can
write
\begin{equation}
  \label{diagT}
  T(\om)=U(\om)\ms D(\om)\,U^\dagger(\om)\,,
\end{equation}
where
\begin{equation}\label{D}
D(\om)=\diag\big(\la_1(\om)\,,\dots,\la_m(\om)\big)
\end{equation}
and $U(\om)$ is the unitary $m\times m$ matrix with entries
\begin{equation}\label{U}
  U_{nn'}(\om)=\frac1{\sqrt m}\,\big[\om\ms\e^{2\pi\iu n'/m}\big]^{m-n}\,.
\end{equation}
For later convenience, we shall also evaluate the sum
\begin{align}
  &\sum_{n,n'=1}^mU_{nk}(\om)\ms\overline{U_{n'k}(\om)}\notag\\
  &=\frac1m\,\sum_{n,n'=1}^m\ms\e^{2\pi\iu k(m-n)/m}
  \e^{-2\pi\iu k(m-n')/m}\notag\\
  &=\frac1m\,\sum_{n,n'=1}^m\e^{2\pi\iu k(n'-n)/m}=m\,\de_{km}\,,
  \qquad |\om|=1\,.\label{Uid}
\end{align}

Let us now go back to the characteristic function~\eqref{charf}. Since
\begin{equation}\label{gaj}
\om_j(\e^{\iu t/\si})=\e^{\iu\ga_jt}\,,\qquad \ga_j\equiv\frac{\cF(j)}{m\si}\,,
\end{equation}
we can apply Eqs.~\eqref{diagT}--\eqref{U} to the matrices $T_j(\e^{\iu t/\si})= T(\e^{\iu
  \ga_jt})$ in Eq.~\eqref{ZT}. We thus readily obtain
\begin{equation}
  \label{charfprod}
  \hat\vp(t)=\frac{\e^{-\iu\mu t/\si}}m\sum_{n,n'=1}^mM_{nn'}(t)\,,
\end{equation}
where the $m\times m$ matrix $M(t)$ is given by
\begin{multline}
  \label{M}
  M(t)=U\big(\e^{\iu\ga_1 t}\big)
  D\big(\e^{\iu\ga_1t})B_1(t)\cdots D\big(\e^{\iu\ga_{N-2}t})B_{N-2}(t)\\
  \times D\big(\e^{\iu\ga_{N-1} t}\big)U^\dagger\big(\e^{\iu\ga_{N-1}t}\big)
\end{multline}
with
\begin{equation}
  \label{Bj}
  B_j(t)=U^\dagger\big(\e^{\iu\ga_j t}\big)U\big(\e^{\iu\ga_{j+1} t}\big).
\end{equation}
{}From Eqs.~\eqref{charfprod}--\eqref{Bj} one can  determine the large $N$
limit of the characteristic function~$\hat\vp(t)$, as we shall next discuss. To this end, we note
first of all that
\begin{equation}\label{gajN}
\ga_j=O(N^{-1/2}).
\end{equation}
Indeed, for the HS and FI chains $\cF(j)$ is $O(N^2)$ by Eq.~\eqref{cF}, while it can be checked
that $\si$ is $O(N^{5/2})$ by substituting the corresponding expressions of $\cF$ in
Eq.~\eqref{cF} into Eq.~\eqref{si2} (cf.~the explicit formulas in Refs.~\cite{FG05} and
\cite{BFGR10}). Likewise, for the PF chain $\cF(j)$ is $O(N)$, while $\si$ is
$O(N^{3/2})$~\cite{BFGR08epl}. By Eq.~\eqref{U}, this implies that
\begin{equation}
  \label{UN}
  U\big(\e^{\iu\ga_j t}\big)=R+O(N^{-1/2})\,,
\end{equation}
where
\begin{equation}
  \label{R}
  R\equiv U(1)
\end{equation}
is a constant unitary matrix (independent of $N$). In order to estimate $B_j(t)$, note first that
for all three chains~\eqref{HS}--\eqref{FI} we have
\begin{equation}
  \label{gadiff}
  \ga_j-\ga_{j+1}=O(N^{-3/2})\,;
\end{equation}
indeed, $\cF(j)-\cF(j+1)$ is $O(N)$ (resp. $O(1)$) for the HS and FI (resp.~PF) chains. Taking
this into account and using Eq.~\eqref{U} we immediately obtain
\begin{align*}
  &\big[B_j(t)\big]_{nn'}=\sum_{k=1}^m
  \overline{U_{kn}\big(\e^{\iu\ga_j t}\big)}U_{kn'}\big(\e^{\iu\ga_{j+1} t}\big)\notag\\
  &= \frac1m\,\sum_{k=1}^m\Big[\e^{-\iu \ga_jt}\e^{-2\pi\iu
    n/m}\Big]^{m-k}\Big[\e^{\iu\ga_{j+1} t}\e^{2\pi\iu n'/m}\Big]^{m-k}\notag\\
  &=\frac1m\,\sum_{k=1}^m\e^{\iu(\ga_{j+1}-\ga_j)(m-k)t}\,\e^{2\pi\iu (n'-n)(m-k)/m}\,,\notag\\
%   &=\frac1m\sum_{k=1}^m\big[1+O(N^{-3/2})\big]\e^{2\pi\iu (n-n')(m-k)/m}\notag\\
  &=\frac1m\,\sum_{k=1}^m\e^{2\pi\iu (n'-n)(m-k)/m}+O(N^{-3/2})\notag\\&=\de_{nn'}+O(N^{-3/2}),
\end{align*}
so that
\begin{equation}
  \label{Bjasympt}
  B_j(t)=\id+O(N^{-3/2})\,.
\end{equation}
On the other hand, from Eq.~\eqref{lak} it easily follows that
\begin{equation}
  \label{las}
  |\la_k(\e^{\iu\ga_j t})|\le1\,,\qquad k=1,\dots,m\,.
\end{equation}
Equation~\eqref{M} and the estimates~\eqref{UN}, \eqref{Bjasympt} and \eqref{las} immediately
yield the asymptotic formula
\begin{equation}
  \label{Masympt}
  M(t)=R\,\La(t)R^\dagger+O(N^{-1/2})\,,
\end{equation}
where $\La(t)$ is the diagonal matrix with entries
\begin{equation}\label{Lak}
  \La_k(t)=\prod_{j=1}^{N-1}\la_k(\e^{\iu\ga_j t})\,,\qquad k=1,\dots, m\,.
\end{equation}
Inserting Eqs.~\eqref{Masympt}-\eqref{Lak} into Eq.~\eqref{charfprod} and using the
identity~\eqref{Uid} with $\om=1$ we  obtain the simple asymptotic estimate
\begin{equation}\label{hatvp}
\hat\vp(t)=\e^{-\iu\mu t/\si}\La_m(t)+O(N^{-1/2})\,.
\end{equation}

In view of the latter equation, in order to complete our proof of Eq.~\eqref{charflim} we just
have to determine the asymptotic behavior as $N\to\infty$ of the eigenvalue $\la_m(\om)$, with
$\om=\e^{\iu \ga_jt}$ unimodular. By Eq.~\eqref{lak}, this eigenvalue is given by
\begin{equation}\label{lam}
  \la_m(\e^{\iu \ga_j t})=\frac1m\sum_{l=0}^{m-1}\e^{\iu l\ga_jt}=\frac{\e^{\iu
      m\ga_jt}-1}{m(\e^{\iu\ga_jt}-1)}
\end{equation}
provided that $t\notin (2\pi/\ga_j)\,\ZZ$. Note that for any fixed $t\ne0$ the condition $mt\notin
(2\pi/\ga_j)\,\ZZ$ is fulfiled for sufficiently large $N$ on account of Eq.~\eqref{gajN}, and
implies the weaker condition $t\notin (2\pi/\ga_j)\,\ZZ$. Thus for all $t\ne0$ Eq.~\eqref{lam}
holds if $N$ is large enough, and $\la_m(\e^{\iu \ga_jt})\ne0$. The latter equation, together with
the elementary Taylor expansion
\[
\log\bigg[\frac{\e^{\iu mx}-1}{m(\e^{\iu x}-1)}\bigg]=\frac12\,(m-1)\ms\iu\ms
x-\frac1{24}(m^2-1)x^2+O(x^4)
\]
and the identity \eqref{mu}, easily yields the asymptotic formula
\begin{multline}
  \label{charffinal}
  -\frac{\iu\mu t}\si+\log\La_m(t)\\=
  -\frac{t^2}{24}\,(m^2-1)\sum_{j=1}^{N-1}\ga_j^2+O(N^{-1})\,.
\end{multline}
In order to estimate the coefficient of $t^2$ in the previous formula, it suffices to note that
Eq.~\eqref{si2} can be equivalently written as
\begin{align*}
  \label{si2asympt}  
  \frac1{12}(m^2-1)\sum_{j=1}^{N-1}\ga_j^2&=1
  +\frac16(m^2-1)\sum_{j=2}^{N-1}(\ga_{j-1}-\ga_j)\ga_j\\
  &= 1+O(N^{-1})\,,
\end{align*}
where we have used Eqs.~\eqref{gajN} and \eqref{gadiff} for the last estimate. Hence
\[
-\frac{\iu\mu t}\si+\log\La_m(t)=-\frac{t^2}2+O(N^{-1})\,,
\]
which obviously implies Eq.~\eqref{charflim} in view of Eq.~\eqref{hatvp}.

We shall conclude by summarizing the main result of this paper and presenting an outline of
related future work. We have rigorously shown that for all spin chains of Haldane--Shastry type
associated with the $A_{N-1}$ root system the level density (normalized to unity) approaches a
Gaussian distribution as the number of sites tends to infinity. Our proof essentially relies on
two key properties of these chains, namely Eq.~\eqref{energies} for the energies in terms of the
motifs~\eqref{dejk}, and the estimates~\eqref{gajN} and~\eqref{gadiff} involving the large
$N$ behavior of the dispersion relation~\eqref{cF}. 

Our results admit several natural generalizations. For instance, one could consider the
$\mathrm{su}(n|n')$ supersymmetric extensions of the chains~\eqref{HS}--\eqref{FI}, some of which
have already been studied in the literature~\cite{BB06,BB09}. It is straightforward to check that
Eq.~\eqref{dejk} for the motifs should be replaced by
\[
\setlength{\abovedisplayskip}{6pt}
\de(j,k)=
\begin{cases}
  1\,,\quad &j>k \enspace\text{or}\enspace j=k>n\\[3pt]
  0\,,& j<k \enspace\text{or}\enspace j=k\le n\,,
\end{cases}\vrule width0pt depth17.75pt
\]
where $j,k=1,\dots,n+n'\equiv m$. As a consequence, the last $n'$ elements in the main diagonal of
the transfer matrix~\eqref{Tom} are replaced by $\om^m$, so that the resulting matrix is no longer
Toeplitz. Although this fact certainly complicates the explicit diagonalization of the transfer
matrix, we believe that the main ideas behind our proof can still be applied to this case.

It is also natural to consider the generalization of our result to spin chains of HS type
associated with other root systems, like $BC_N$ or $D_N$. The main difficulty in this respect is
the fact that for these chains no description of the energies in terms of motifs akin to
Eq.~\eqref{energies} is known so far. At least for the Sutherland spin chain of $BC_N$
type~\cite{BPS95, EFGR05}, some preliminary results of our group indicate that such a description
is possible, and that our proof can be suitably adapted to this case.

\begin{acknowledgments}
This work was supported in part by the MICINN and the UCM--Banco Santander
under grants no.~FIS2008-00209 and~GR58/08-910556. The authors would also like to thank
Gabriel \'Alvarez for some useful discussions.
\end{acknowledgments}
%
% \bibliography{apsrefs}
% \bibliographystyle{apsrev}
% 

\end{document}